\documentclass{webofc}

\usepackage[varg]{txfonts}   
\usepackage{hyperref}
\usepackage{url}
\hypersetup{colorlinks=true,citecolor=blue,urlcolor=blue,linkcolor=blue}
\usepackage{slashed}

\DeclareMathOperator{\Tr}{\rm Tr}
\begin{document}
\title{QCD phase structure at finite isospin chemical potential and smaller-than-physical quark mass}

\author{\firstname{Bastian~B.}~\lastname{Brandt}\inst{1}\fnsep\thanks{\email{brandt@physik.uni-bielefeld.de}} \and
        \firstname{Volodymyr}~\lastname{Chelnokov}\inst{2}\fnsep\thanks{\email{chelnokov@itp.uni-frankfurt.de}} \and
        \firstname{Francesca}~\lastname{Cuteri}\inst{2}\fnsep\thanks{\email{cuteri@itp.uni-frankfurt.de}}
        \and
        \firstname{Gergely}~\lastname{Endr\H{o}di}\inst{3}\fnsep\thanks{\email{gergely.endrodi@ttk.elte.hu}}
}

\institute{
Institute for Theoretical Physics, University of Bielefeld, D-33615 Bielefeld, Germany
\and
Institut f\"{u}r Theoretische Physik, Goethe-Universit\"{a}t Frankfurt, \\
 Max-von-Laue-Str.\ 1, 60438 Frankfurt am Main, Germany 
\and
Institute of Physics and Astronomy, Eötvös Lor\'and University, \\ P\'azm\'any P. s\'et\'any 1/A, H-1117 Budapest, Hungary
}

\abstract{Introduction of a nonzero isospin chemical potential in QCD leads to the emergence of a pion condensed phase at sufficiently large $\mu_I$, bounded by a second order transition line. At zero temperature the pion condensate appears at $\mu_I = m_\pi / 2$. Recent numerical studies at physical quark masses show that the pion condensation boundary remains vertical up to the meeting point with the chiral crossover line. If this result remains valid when the light quark mass (and the pion mass) goes to zero, then in the chiral limit at temperatures below the chiral transition pion condensation happens at arbitrary nonzero $\mu_I$. We report on results of a lattice QCD simulation of a 2+1 flavour QCD at nonzero isospin chemical potential, at smaller-than-physical light quark mass, that support this scenario.
}
\maketitle
\section{Introduction}
\label{intro}
Most physical strongly interacting systems exhibit isospin asymmetry -- an imbalance in the number of up and down quarks. In QCD, isospin asymmetry is implemented via an isospin chemical potential, $\mu_I = (\mu_u - \mu_d) / 2$, in addition to the baryon chemical potential, $\mu_B = (\mu_u + \mu_d) / 2$. Introduction of the isospin chemical potential naturally extends the QCD phase diagram into three dimensions with coordinates $(T,\  \mu_B,\ \mu_I)$. 
The theory is sign-problem-free at $\mu_B = 0$ and arbitrary value of $\mu_I$ 
\cite{finite-isospin}, which allows direct Monte Carlo simulations. This makes isospin-asymmetric QCD (at zero $\mu_B$) a valuable testbed for studying the properties of the methods designed to tackle the sign problem \cite{taylor-reliability}, as well as a useful region to start the extrapolation into physically interesting nonzero $\mu_B$ values \cite{isospin-reweighting,isospin-taylor}. 

\begin{figure*}[h]
\centering
\includegraphics[width=6cm,clip]{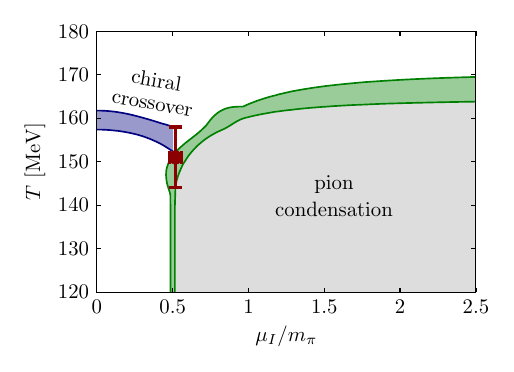}
\includegraphics[width=6cm,clip]{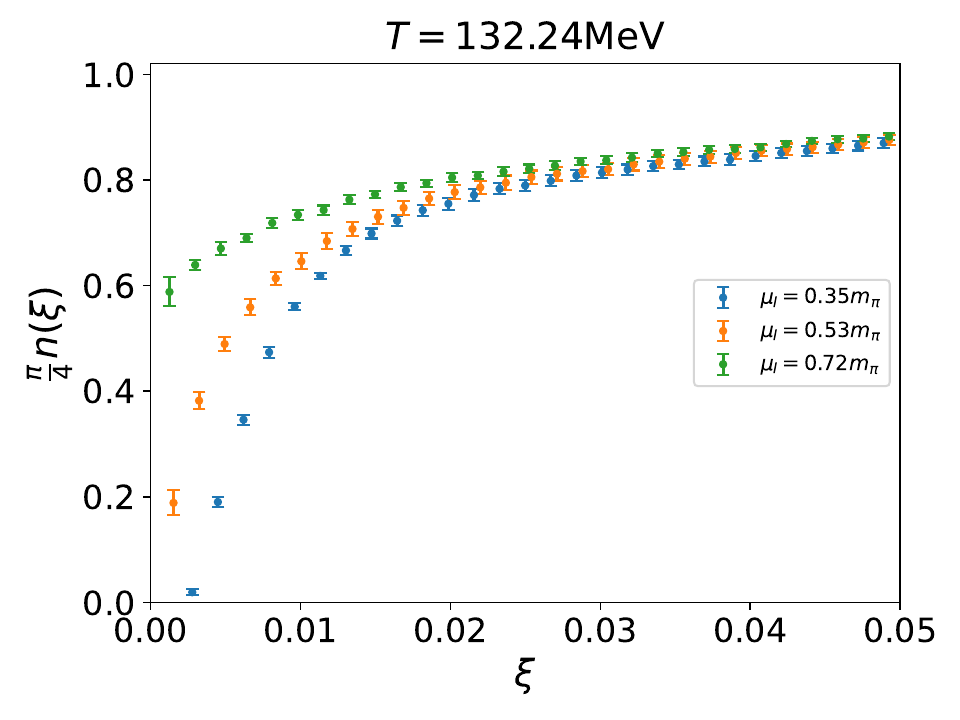}
\caption{
(left) Pion condensation phase at physical light quark mass (taken from \cite{muI-phys-plot}).
(right) Singular value densities at $T=132.24$ MeV, for three different values of $\mu_I$.}
\label{improved_condensate_densities} 
\end{figure*}

The QCD phase structure in the $T$ -- $\mu_I$ plane features a pion condensed phase, characterized by a Bose–Einstein condensate of charged pions~\cite{muI-phys, muI-phys-plot} (see figure~\ref{improved_condensate_densities}~(left)). Effective models predict a more complex phase structure in the full $T$ -- $\mu_B$ -- $\mu_I$ plane,  e.g. \cite{andersen-noetvedt}.

Since at low temperatures the pion condensation line remains vertical at $\mu_I = m_\pi / 2$, and the pion mass $m_\pi$ vanishes as $\sqrt{m_{ud}}$ in the chiral limit, it is natural to expect that this picture remains valid also at smaller-than-physical light quark masses, resulting in appearance of a pion condensate at arbitrary $\mu_I > 0$ in the chiral limit. Our recent study using a renormalization group invariant mean-field model supports this conclusion, while staying in good agreement with numerical data \cite{rg-invariant-model}. 
In this work, we perform a direct lattice Monte Carlo study of the pion condensation phase boundary at lighter-than-physical quark mass, in order to test the chiral scenario described above.

\section{Lattice simulation setup} 

We follow the same lattice setup as in~\cite{muI-phys}, with the light quark mass set to $m_{ud} = m_{ud, phys}/2$. The light fermion determinant, after simplifications, takes the form
\begin{equation}
{\rm det}\;\mathcal{M}_{ud} = {\rm det} \left( \left| \slashed{D}(\mu_I) + m_{ud}\right|^2 + \lambda^2 \right) \ , 
\end{equation}
where $\lambda$ is an unphysical pion source term, explicitly breaking the residual $U(1)$ symmetry for $\mu_I \neq 0$ and effectively fixing the direction of the pion condensate in the $\lambda \to 0$ limit.
The $\lambda$ term also serves as a regulator for simulations, increasing each singular value of the fermion operator by $\lambda^2$, making the numeric inversion problem better conditioned.
The pion condensate can be obtained by differentiating the partition function with respect to $\lambda$.
\begin{equation}
     \pi^\pm  = 
    \frac{T}{2 V} \Tr \frac{\lambda}{\left| \slashed{D}(\mu_I) + m_{ud}\right|^2 + \lambda^2} \ .
\end{equation}
Since $\lambda$ is an unphysical term, we perform the simulations at several nonzero $\lambda$ values and then 
perform an extrapolation to $\lambda = 0$. 
To improve the 
extrapolation we employ the Banks–Casher type relation for the pion condensate described in~\cite{muI-phys}
\begin{equation}
    \left\langle \pi^\pm \right\rangle = 
    \frac{\pi}{4} \left\langle \rho(0) \right\rangle\ .
\end{equation}
Here, $\pi_\pm$ is the pion condensate, and $\rho(0)$ is the density of the fermion operator singular values at zero. In practice we extract $\rho(0)$ from the $k = 150$ smallest singular values of the fermion operator (with $\lambda = 0$) measured on configurations generated at finite $\lambda$, extrapolating the densities $n(\xi) = \frac{1}{\xi}\int_0^\xi \rho(\eta) d \eta$ to $\xi = 0$. 
The behavior of the density for different phases is shown in figure~\ref{improved_condensate_densities}~(right). The phase transition point is identified as the value of $\mu_I$ at which $\lim_{\xi \to 0} n(\xi)$ vanishes.

Simulations are performed with 2+1 flavors of staggered fermions, with a tree-level Symanzik-improved gauge action. Most simulations are done on $24^3 \times 8$ lattices, with several points recalculated
on $32^3 \times 10$ and $36^3 \times 12$ lattices to check the effect of finite lattice spacing on the results. 
At each temperature (4 different temperatures between 114 MeV and 142 MeV) we perform a scan in $\mu_I$, taking 5-7  $\mu_I$ values per one temperature value. Similarly, we perform a scan in temperature at a fixed $\mu_I$, with 9 different values of temperature, to extract the transition temperature corresponding to $\mu_I \approx 0.72 m_\pi$.
For each set of parameters, 3 values of the pion source $\lambda$ were simulated to enable extrapolation to zero pion source.
Each measurement is done with at least 200 configurations, following 1000 thermalization updates. 

\section{Reweighting}

To obtain explicit observable dependence on the pion source $\lambda$ and thus improve the extrapolation to $\lambda = 0$ we perform reweighting of the gauge field configurations from the $\lambda_{\mathrm{sim}}$ at which they were sampled, to a nearby value $\lambda_\mathrm{new}$, introducing weights
\begin{equation}
W(\lambda_\mathrm{sim};\lambda_\mathrm{new}) = \frac{\left(\det \left[ |\slashed{D}(\mu_I) + m_{ud}|^2 +\lambda_\mathrm{new}^2 \right] \right)^{1/4}}{\left(\det \left[|\slashed{D}(\mu_I) + m_{ud}|^2 + \lambda_\mathrm{sim}^2\right]\right)^{1/4}}\ .
\label{exact-reweighting}
\end{equation}

Since calculating the exact determinant ratio is computationally expensive, and does not yield an explicit formula for the $\lambda$ dependence, we can approximate the ratio with at leading order by
\begin{equation}
\log W_\mathrm{LO}(\lambda_\mathrm{sim},\lambda_\mathrm{new}) \approx 
- \frac{\lambda_\mathrm{sim}^2-\lambda_\mathrm{new}^2}{4} \Tr \frac{1}{|\slashed{D}(\mu_I) + m_{ud}|^2 + \lambda_\mathrm{sim}^2} = 
- \frac{\lambda_\mathrm{sim}^2-\lambda_\mathrm{new}^2}{\lambda_\mathrm{sim}} \frac{V}{ 2 T} \pi^\pm \ ,
\label{leading-order-reweighting}
\end{equation}

This approximation is further improved by using exact contribution of the smallest $k$ singular values (that are calculated for density extraction):
\begin{equation}
\log W(\lambda_\mathrm{sim},\lambda_\mathrm{new})  \approx \log W_{LO}(\lambda_\mathrm{sim},\lambda_\mathrm{new}) + \frac{1}{4} \sum_{i=1}^{k} \left( \log \frac{\xi_i^2+ \lambda_\mathrm{new}^2}{\xi_i^2 + \lambda_\mathrm{sim}^2} +  \frac{\lambda_\mathrm{sim}^2- \lambda_\mathrm{new}^2}{\xi_i^2 + \lambda_\mathrm{sim}^2} \right) \ ,
 \label{beyond-leading-order-reweighting}
\end{equation}

To fully use our data at three different $\lambda$ values, we employ multihistogram reweighting~\cite{Ferrenberg-Swendsen} combining ensembles simulated at different $\lambda$. 
We restrict reweighting to the range between the largest and smallest simulated $\lambda$, followed by a final linear extrapolation to $\lambda=0$.

\begin{figure*}[h]
\centering
\includegraphics[width=6cm,clip]{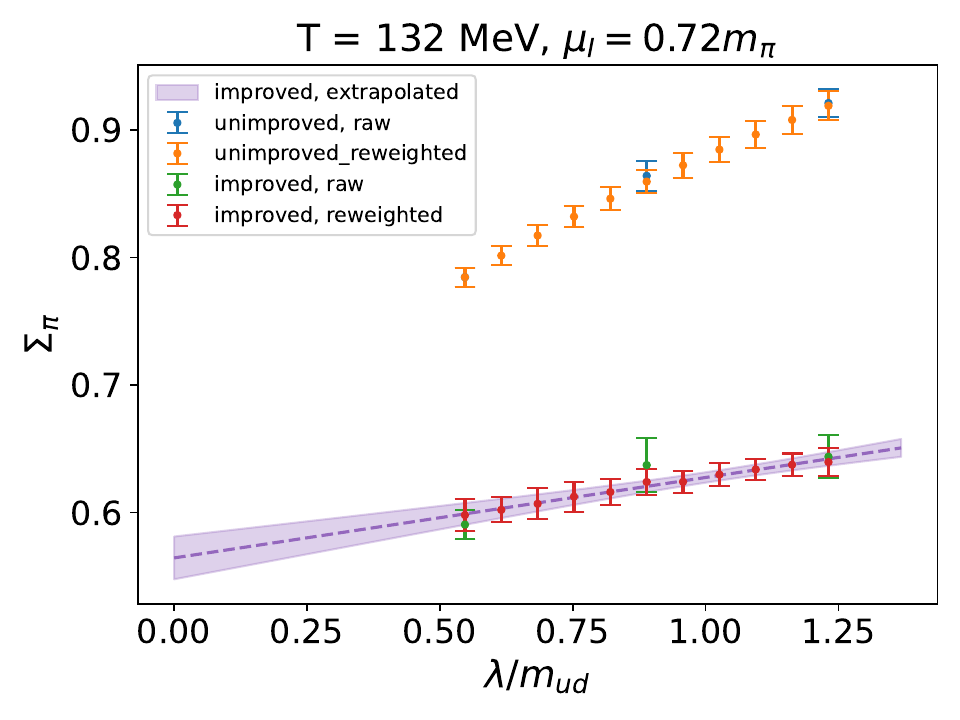}
\includegraphics[width=6cm,clip]{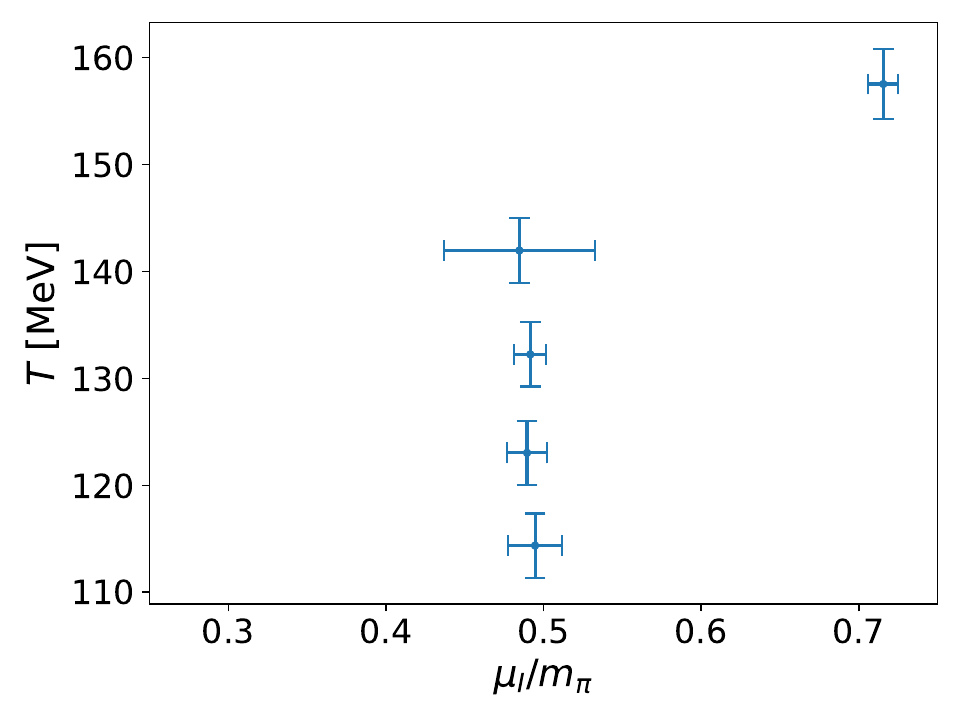}
\caption{(left)
Reweighting of the pion condensate in pion source $\lambda$ and final extrapolation of the reweighted improved condensate to $\lambda=0$.
(right) The pion condensation boundary points obtained at $m_\pi = m_{\pi,\mathrm{phys}} / \sqrt{2}$.}
\label{pion-condensation-boundary} 
\end{figure*}

An example of the procedure is shown in figure~\ref{pion-condensation-boundary}~(left), together with results for the unimproved pion condensate and its reweighting. The collected pion condensation boundary points are shown in figure~\ref{pion-condensation-boundary}~(right).

\section{Summary}
We determined the pion condensation boundary at $m_\pi = m_{\pi,\mathrm{phys}} / \sqrt(2)$ from direct Monte Carlo simulations. The boundary remains vertical up to $T = 142\  \text{MeV}$, supporting the scenario of pion condensation that pion condensation occurs at arbitrarily small $\mu_I$ for $0 < T < T_c$ in the chiral limit, as predicted by effective models. This implies that in the chiral limit (at least for $\mu_B = 0$), the hadronic matter phase exists only along the line $\mu_I = 0$, and that the chiral transition temperature $T_c$ corresponds to a multicritical point.

\section*{Acknowledgments}

This work was supported by the Deutsche Forschungsgemeinschaft (DFG, German Research Foundation) – project number 315477589 – TRR 211. GE acknowledges support by the Hungarian National Research, Development and Innovation Office (Research Grant Hungary 150241) and the European Research Council (Consolidator Grant 101125637 CoStaMM). The authors acknowledge the use of the Goethe-HLR and Bielefeld clusters and thank the computing staff for their support.


\begin{thebibliography}{}
%
%


\bibitem{finite-isospin}
D.~Son, M.~Stephanov, 
QCD at finite isospin density.
Phys.Rev.Lett. \textbf{86}, 592 (2001).
\url{https://doi.org/10.1103/PhysRevLett.86.592}

\bibitem{taylor-reliability}
B.~Brandt, G.~Endr\H{o}di,
Reliability of Taylor expansions in QCD.
Phys.Rev.D \textbf{99}, 014518 (2019).
\url{https://doi.org/10.1103/PhysRevD.99.014518}

\bibitem{isospin-reweighting}
B.~Brandt, F.~Cuteri, G.~Endr\H{o}di, S.~Schmalzbauer,
Exploring the QCD phase diagram via reweighting from isospin chemical potential.
PoS \textbf{LATTICE2019}, 189 (2019).
\url{https://doi.org/10.22323/1.363.0189}

\bibitem{isospin-taylor}
B.~Brandt, G.~Endr\H{o}di, G.~Mark\'o,
Exploring the QCD phase diagram via reweighting from isospin chemical potential.
PoS \textbf{LATTICE2019}, 189 (2019).
\url{https://doi.org/10.22323/1.363.0189}

\bibitem{muI-phys}
B.~Brandt, G.~Endr\H{o}di, S.~Schmalzbauer,
QCD phase diagram for nonzero isospin-asymmetry.
Phys.Rev.D \textbf{97}, 054514 (2018).
\url{https://doi.org/10.1103/PhysRevD.97.054514}

\bibitem{muI-phys-plot}
B.~Brandt, G.~Endr\H{o}di, S.~Schmalzbauer,
QCD at nonzero isospin asymmetry.
PoS \textbf{Confinement2018}, 260 (2019).
\url{https://doi.org/10.22323/1.336.0260}

\bibitem{andersen-noetvedt}
J.~Andersen, M.~Nødtvedt,
Pion condensation versus 2SC, speed of sound, and charge neutrality effects in the quark-meson diquark model.
[arXiv:hep-ph/2502.04025] (2025).
\url{https://doi.org/10.48550/arXiv.2502.04025}

\bibitem{rg-invariant-model}
B.~Brandt, V.~Chelnokov, G.~Endr\H{o}di, G.~Mark\'o, D.~ Scheid, L.~von~Smekal, 
Renormalization group invariant mean-field model for QCD at finite isospin density.
Phys.Rev.D \textbf{112}, 054038 (2025).
\url{https://doi.org/10.1103/fryz-f3vw}

\bibitem{Ferrenberg-Swendsen}
A.~Ferrenberg, R.~Swendsen,
Optimized Monte Carlo analysis.
Phys.Rev.Lett. \textbf{63}, 1195 (1989).
\url{https://doi.org/10.1103/PhysRevLett.63.1195}

\end{thebibliography}
\end{document}